\documentclass{elsart1p}
\usepackage{graphicx}
\usepackage{amssymb}

\newcommand{\pip}{$\pi^{+}$}
\newcommand{\pim}{$\pi^{-}$}
\newcommand{\kap}{K$^{+}$}
\newcommand{\kam}{K$^{-}$}
\newcommand{\pbar}{$\rm\overline{p}$}

\newcommand{\s}{$\sqrt{s}$}
\newcommand{\pt}{\ensuremath{p_{\rm t}}}
\newcommand{\dedx}{d$E$/d$x$}

\newcommand{\Gc}{GeV/$c$}

\begin{document}
\begin{frontmatter}
%
%
%
%
%
\title{Strange Particle Production in pp Collisions\\ at \s\ =~0.9 and 7 TeV
measured\\ with the ALICE Experiment}
%
%

\author{Helmut Oeschler, for the ALICE Collaboration}

\address{Darmstadt University of Technology, Germany
and CERN, Geneva, Switzerland}

\begin{abstract}

Hadrons measured in  proton-proton collisions at \s~ = 0.9 and 7
TeV with the ALICE detector have been identified using various
techniques: the specific energy loss and the time-of flight
information for charged pions, kaons and protons, the displaced
vertex resulting from their weak decay for K$^{0}$, $\Lambda$ and
$\Xi$ and the kink topology of decaying charged kaons. These
various particle identification tools give the best separation at
different momentum ranges and the results are combined to obtain
spectra from $\pt = 100$ MeV/$c$ to 2.5 GeV/$c$. This allows to
extract total yields. In detail we discuss the K/$\pi$ ratio
together with previous measurements and we show a fit using a
statistical approach.
\end{abstract}

\end{frontmatter}

\section{Introduction}
\label{i}

The bulk of particles produced both in heavy-ion and in pp
collisions  has transverse momenta below 1 \Gc. Their spectral
shapes and their composition are major observables that help
understanding the hadronic interactions and the hadronization
process. With the Large Hadron Collider (LHC) at CERN much higher
energies are reached opening a new area where particle production
occurs mainly via hard interactions.

From about 1 $A$~GeV up to the highest energies a statistical
concept has successfully described the chemical composition of
hadrons, consisting  of u, d and s quarks, measured in heavy-ion
reactions and to some extend also in elementary collisions. The
much higher collision energies at the LHC will crucially test this
picture in the realms of hard partonic interactions. Furthermore,
it will allow to compare results from pp collisions at high
multiplicities with those from the heavy-ion collisions at the
same multiplicity.

The ALICE detector with its excellent particle identification
(PID) capability is perfectly suited for these studies. In these
proceedings the various PID techniques are shortly mentioned and
the results obtained in pp collisions at \s\ = 0.9 TeV are
presented.

\section{The ALICE Experiment and data analysis}

The ALICE detector and its performance are described in detail in
\cite{Alessandro:2006yt,Carminati:2004fp,ALICE-JINST}. The first
pp collisions at \s\ =~0.9 TeV are analysed using various PID
techniques and the results are reported in two recent
papers~\cite{strange,spectra}.

Charged $\pi$, K and p are identified using the Inner Tracking
System (ITS), the Time Projection Chamber (TPC) and the
Time-Of-Flight array (TOF)~\cite{spectra}.
Weakly decaying strange particles are identified by their
displaced vertex using the tracking information in ITS and TPC.
These studies are reported in detail in~\cite{strange}. Charged
kaons decaying within the TPC are identified by their typical kink
topology. All three techniques for kaon identification agree very
well and have optimal performance at different
momenta~\cite{strange,spectra}.

\begin{figure}[h]
\centering
\resizebox{0.49\textwidth}{!}{%
\includegraphics{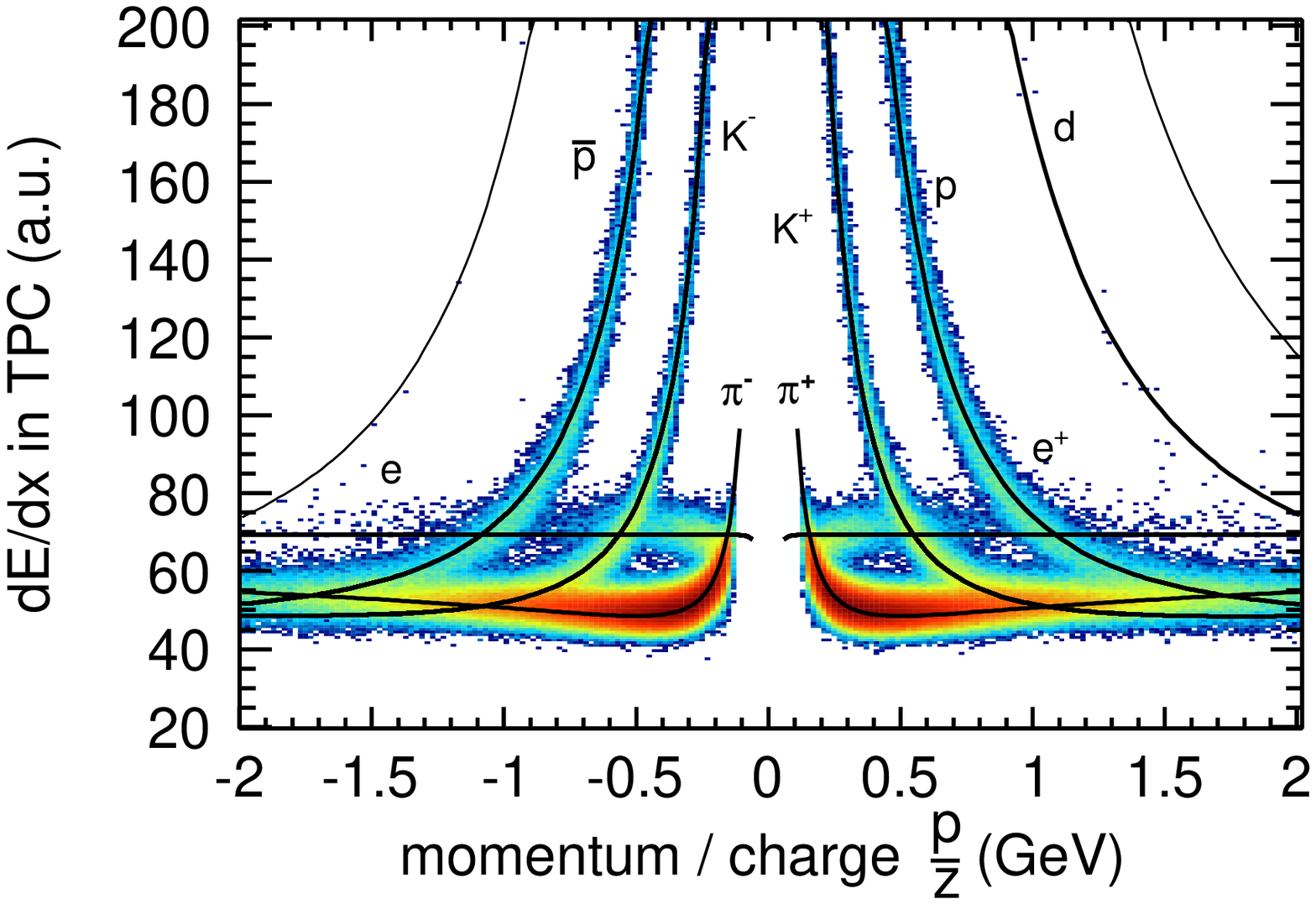}
}
\resizebox{0.49\textwidth}{!}{%
\includegraphics{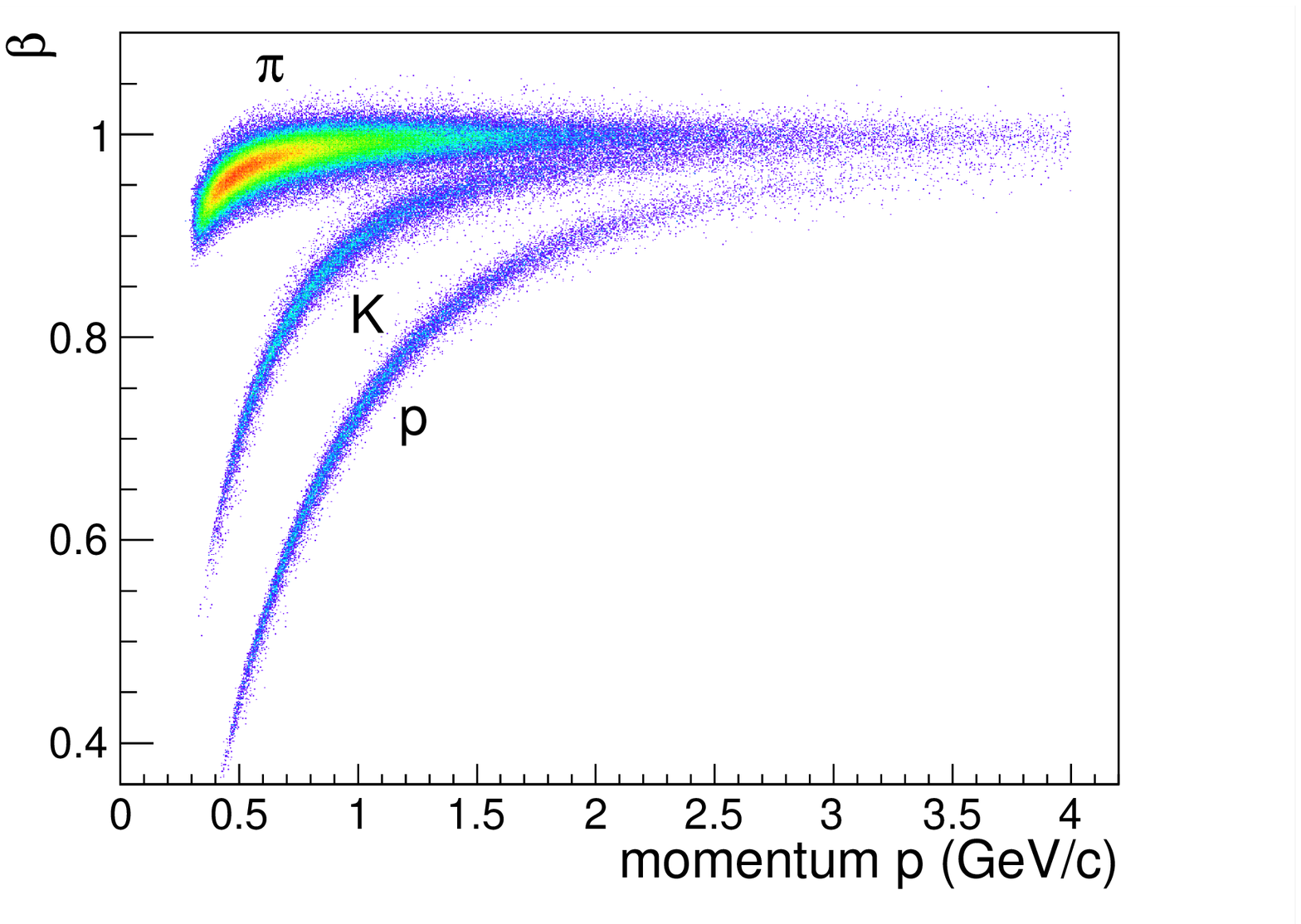}
} \caption{Examples of PID techniques using in pp collisions at
0.9 TeV~\cite{spectra}. Left: Specific energy loss \dedx\
vs.~momentum for tracks measured with the ALICE TPC. The solid
lines are a parametrization of the Bethe-Bloch curve. Right:
$\beta$ of particles measured by TOF vs. their momentum.}
\label{ana}
\end{figure}

Particle identification in the TPC is based on the specific energy
deposit by each particle in the drift gas, shown in
Fig.~\ref{ana}, left, as a function of momentum separately for
positive and negative charges.   The solid curves show Bethe-Bloch
parametrization.
The ITS being closer to the interaction point, is able to identify
also particles with low momenta which are absorbed before reaching
the TPC using the \dedx\ in the silicon material.
Particles reaching the TOF system are identified by measuring
their momentum and velocity simultaneously. The velocity $\beta=
L/t_{\rm TOF}$ is obtained from the measured time of flight
$t_{\rm{TOF}}$ and the reconstructed flight path $L$ along the
track trajectory.
The measured velocities are shown in Fig.~\ref{ana}, right, as a
function of the momentum $p$ at the vertex. The bands
corresponding to charged pions, kaons and protons, are clearly
visible and show that particles can be separated up to about 2.5
\Gc.

\section{Results}

\begin{figure}[h]
\centering
\resizebox{0.49\textwidth}{!}{%
\includegraphics{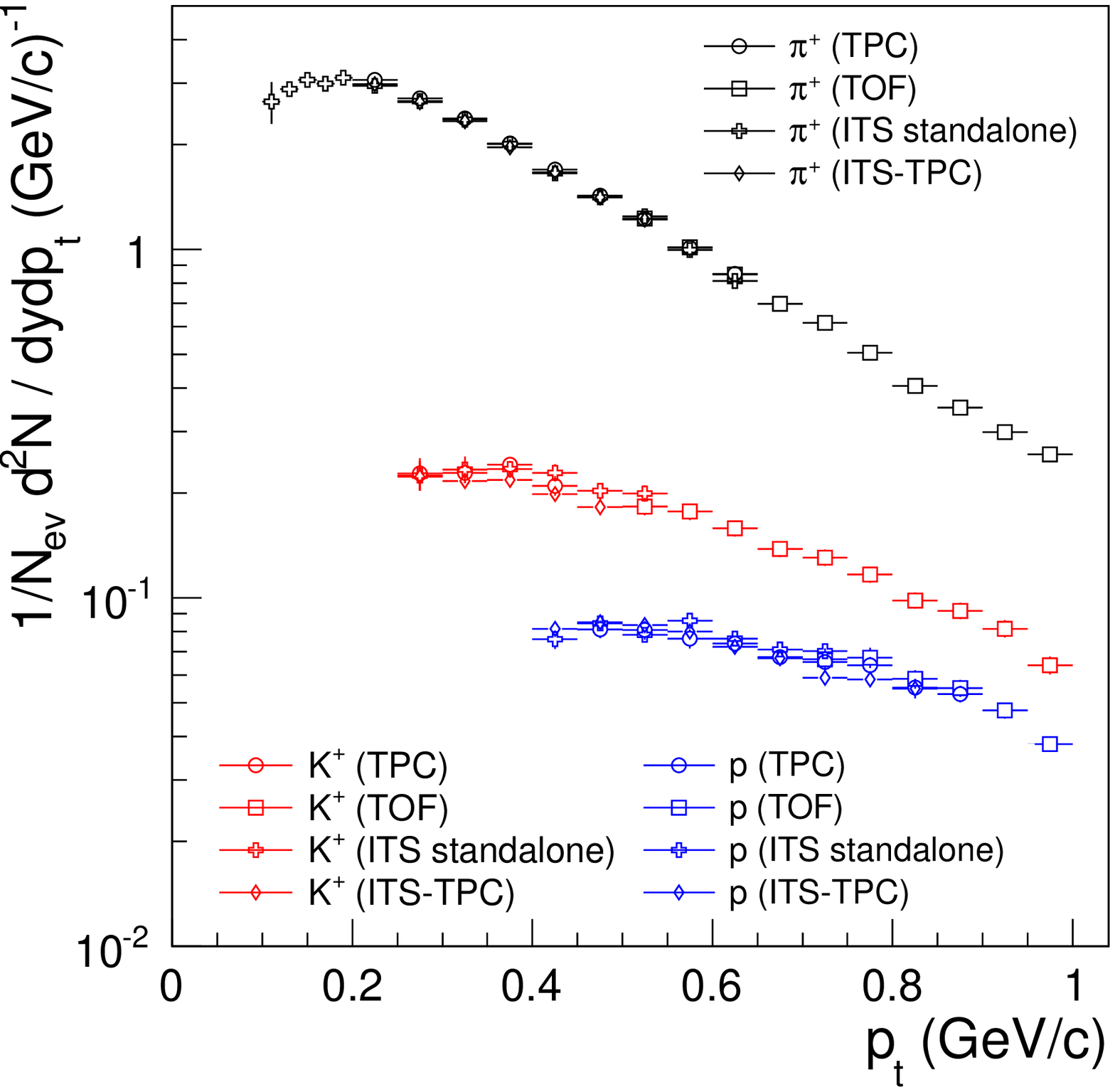}
}
\resizebox{0.49\textwidth}{!}{%
\includegraphics{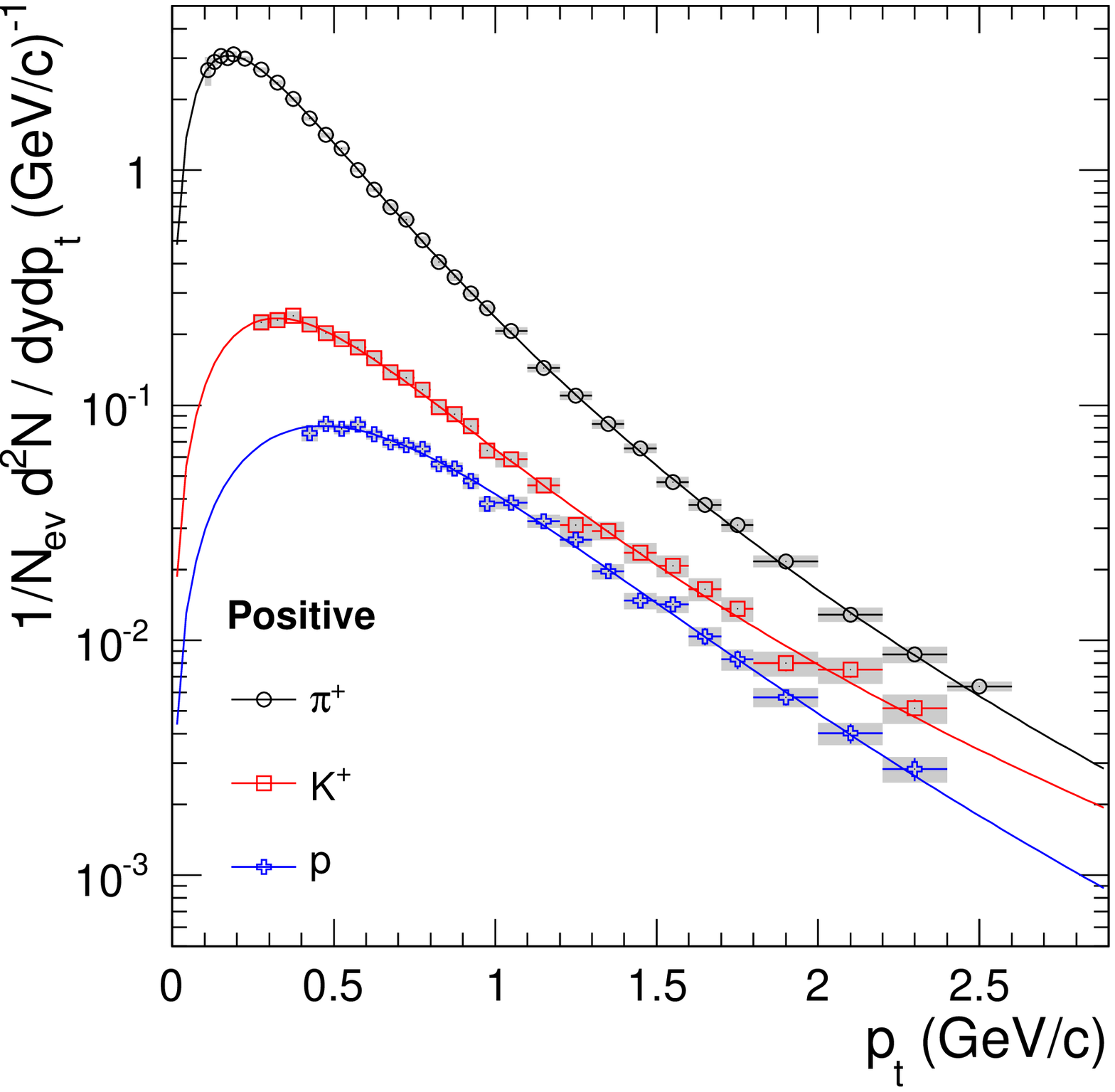}
} \caption{Left: Transverse momentum spectra ${\rm d}^2N / ({\rm
d}p_{\rm t}{\rm d}y)$
 for $|y|<$ 0.5 of positive  hadrons from the various analyses.
 Only systematic errors are plotted. Right: Transverse momentum spectra of positive
 hadrons from pp collisions at \s\ = 900
GeV. Grey bands: total \pt-dependent error (systematic plus
statistical); normalization systematic error (3.6\%) not shown.
The curves represent fits using a L\'{e}vy function. Both
from~\cite{spectra}.} \label{spec}
\end{figure}

Figure~\ref{spec} shows the spectra of positive pions, kaons and
protons using different PID techniques: ITS stand-alone, ITS-TPC
combined, TPC and TOF. Each of them is optimal for a given
momentum range. The good agreement demonstrates that the relevant
efficiencies are well reproduced by the detector simulations.

The spectra from ITS standalone, TPC and TOF are combined in order
to cover a large momentum range as shown in Fig.~\ref{spec},
right. The spectra have been averaged, using the systematic errors
as weights.  From this weighted average, the combined,
\pt-dependent, systematic error is
derived. 
These spectra are fitted with L\'{e}vy (or Tsallis)
functions~\cite{spectra}.
 This function gives a good description of the
spectra and has been used to extract the total yields and the
$\langle p_{\rm t} \rangle$. The fraction of the yield contained
in the extrapolation of the spectra to zero momentum ranges from
10\% to 25\%. The extrapolation to infinite momentum gives a
negligible contribution.

\begin{figure}[h]
\centering
\resizebox{0.49\textwidth}{!}{%
\includegraphics{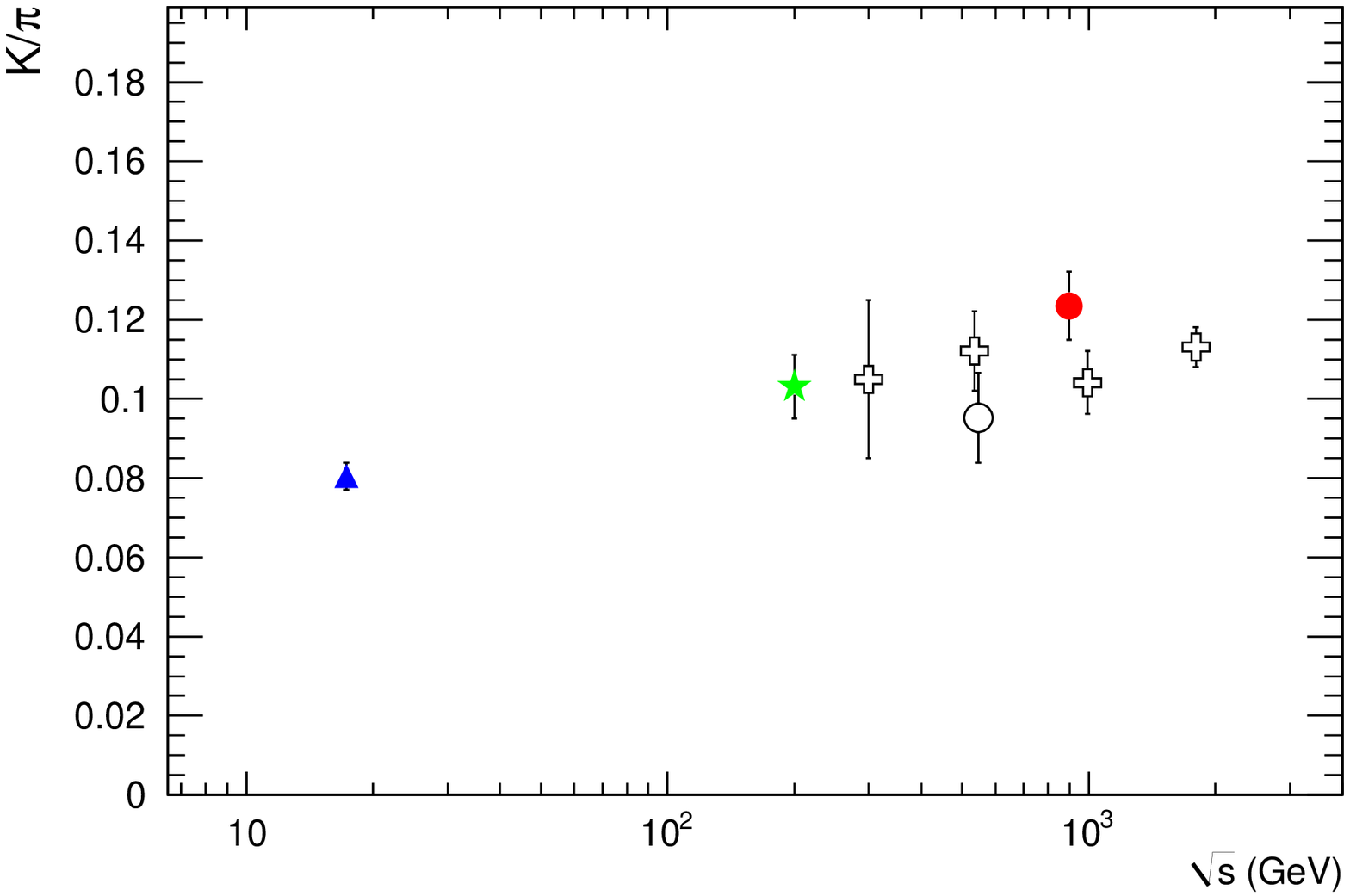}
}
\resizebox{0.49\textwidth}{!}{%
\includegraphics{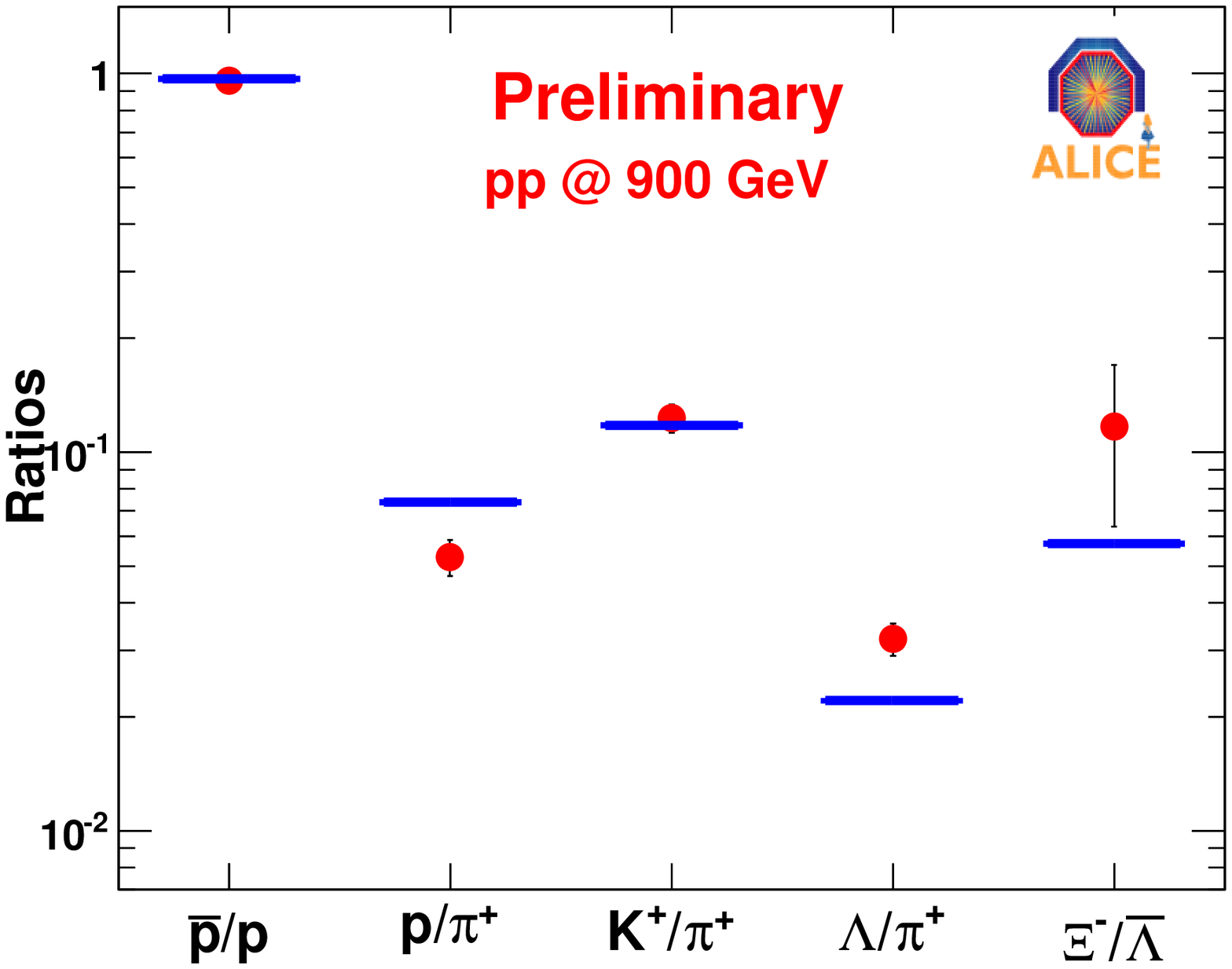}
} \caption{Left: Ratios (\kap+\kam)/(\pip+\pim) and K$^0/\pi$ as a
function of \s. The solid red point refers to the ALICE result.
Other data are from pp
collisions
(full symbols, NA49, STAR) and from \pbar p interaction
(open symbols, E735, UA5). For references see~\cite{spectra}.
Right: Fit of the measured particle ratios with the statistical
model code THERMUS exhibiting a rather poor description.}
\label{str}
\end{figure}

In Fig.~\ref{str}, left, we show the K/$\pi$ ratio as a function
of \s\ both in pp (full symbols)
and in \pbar p (open symbols)
collisions. The ALICE result is the solid red point. For most
energies, (\kap+\kam)/(\pip+\pim) is plotted, but for some cases
K$^{0}/\pi^0$ is used instead. This ratio measured in pp reactions
varies from \s = 200 GeV (K/$\pi$ =
$0.103\pm0.008$)~\cite{:2008ez} to \s = 900 GeV
(K/$\pi$=$0.123\pm0.004\pm0.010$), yet consistent within the error
bars. The results at 7 TeV will show whether the K/$\pi$ ratio
keeps rising slowly as
 a function of \s{} or whether it saturates.

The measured yields of charged $\pi$, K, p and of K$^0$, $\Lambda$
and $\Xi$ \cite{strange,spectra} together with the measured \pbar
/p ratio \cite{Aamodt:2010dx} have been used in a
statistical-model fit with THERMUS~\cite{Wheaton:2004qb}. With a
canonical description for the strangeness, we obtain a rather poor
fit with a moderate $\chi^2/{\rm ndf}$ of 7 as can be seen in
Fig.~\ref{str}, right. The fit overestimates the proton yield by
about 30\% while both the $\Lambda$ and $\Xi$ yields are
underestimated.
In view of this unsatisfying description, the resulting thermal
parameters have to be taken with caution. We obtain  a freeze-out
temperature $T$ of $161 \pm 4$ MeV, a baryo-chemical potential
$\mu_B$ of $3\pm2$ MeV and a canonical radius $R_c$ of
$1.35\pm0.07$ fm where the radius of the fireball and the
strangeness correlation volume are kept equal and $\gamma_s = 1$.
As the K/$\pi$ ratio is well described, fitting in addition
$\gamma_s$ results in very similar parameters. The $\phi$ has not
been included in the fit. The inclusion of the $\phi$ meson causes
only minor changes.
It is interesting to note that the thermal fits to the STAR pp
data~\cite{Abelev:2006cs} gave a better description, however
similar deviation of the p and $\Lambda$ yields can be noticed,
yet within the errors.

The results from the pp data at \s\ = 7 TeV will be very important
to observe whether the K/$\pi$ ratio will increase with \s. These
data will then allow a detailed analysis using high-multiplicity
events.
\bibliographystyle{elsarticle-num}
\bibliography{id_goa}




\end{document}